\documentclass{ws-procs961x669}            

\usepackage{amsmath}
\begin{document}
\title{Time-reparametrization invariance: from  Glasses to toy Black Holes}

\author{Jorge Kurchan}

\address{Laboratoire de Physique de l’\'Ecole Normale Sup\'erieure,\\
 ENS, Universit\'e PSL, CNRS, Sorbonne Universit\'e,
Universit\'e de Paris,\\
$^*$E-mail:\\
jorge.kurchan$@$phys.ens.fr
}

\begin{abstract}
Glassy  dynamics have time-reparametrization `softness': glasses  fluctuate, and respond  to external perturbations, primarily by changing  the pace of their  evolution. Remarkably, the same situation also appears in toy models of quantum field theory such as the Sachdev-Ye-Kitaev (SYK) model, where the excitations associated to reparametrizations play the role of an emerging `gravity'. I describe here how these two seemingly unrelated systems share  common features, arising from a technically very similar  origin.   This connection is particularly close between glassy dynamics and supersymmetric variants of the SYK model,
which I discuss  in some detail.  Apart from the  curiosity that this correspondence naturally arouses, there is also the hope that developments in  each field may be useful for the other.
\end{abstract}


\bodymatter

\section{Introduction } \label{intro}

Glasses are, by definition, systems that take very long - perhaps infinite - time to reach thermal equilibrium. It is natural then to try to understand the nature of  such  dynamics:  is there something simple and generic that one can say about this physical situation, at least for the asymptotic limit in which relaxation becomes very slow, yet still far from equilibrium?

In this paper I concentrate on one such feature, that holds exactly in some
solvable models, and at least to a good approximation in real systems: they  develop an emergent time-reparametrization invariance: like a film that remains the same, but the speed at which it is  projected is very weakly determined. Physically this means two things: {\it i)} small external perturbations may drastically alter the speed of evolution, and {\it ii)} spontaneous fluctuations of the `speed' become large.

The first to note that reparametrization invariance emerged from a mean-field dynamical model were (to my knowledge)
Sompolinsky and Zippelius 1982 \cite{Sompolinksy1982}. Years later, the problem arose in the analytic
solution of relaxation \cite{Cugliandolo_1993_Analytical}, mostly as a nuisance. Over the years, it was recognized that reparametrization softness explained the dramatic dynamical effect of driving a glass, for example with shear \cite{cugliandolo1997glassy}. Later, in a remarkable series of papers, \cite{Castillo2002,Castillo2003,Chamon2002,Chamon2007}, the spontaneous fluctuations of time-reparametrizations were studied, and a `sigma model' for reparametrizations was proposed.  
Even the consistency of the Parisi ansatz for a finite-dimensional systems was recently argued to require  the  existence of reparametrization softness \cite{kurchan2023time}.

On a separate development, the quantum dynamics of certain fermion (Sachdev-Ye-Kitaev \cite{SachdevYe1993,KitaevLectures})  models were shown to develop 
the same kind of invariance, and for exactly the same technical reasons, at low temperatures, when
the imaginary time is large, and the dynamics become slow \cite{SachdevYe1993,Parcollet1999,Georges2000,Georges2001,KitaevLectures,Maldacena2016SYK,Polchinski2016}.
Reparametrizations embody the low energy properties of the system, and it was noted by Kitaev \cite{KitaevLectures} that concentrating on these, one obtains a `toy' version of how gravity and Black Holes emerge from
a quantum field theory, saturating the expected bounds on chaos. Quite naturally, a sigma model of reparametrizations was developed \cite{Maldacena2016SYK} representing
this. 

That two rather similar routes would lead to an insight in two  very different contexts 
inevitably arouses curiosity, and, from a practical point of view, suggests that technical tools could be exchanged. A preliminary exploration of this striking connection has been presented in Ref.  \cite{facoetti2019classical}.

\section{Glassy dynamics, classical or quantum} \label{dos}

\subsection{Generics}
\begin{figure}
\hspace{.5in}\includegraphics[width=3.4in]{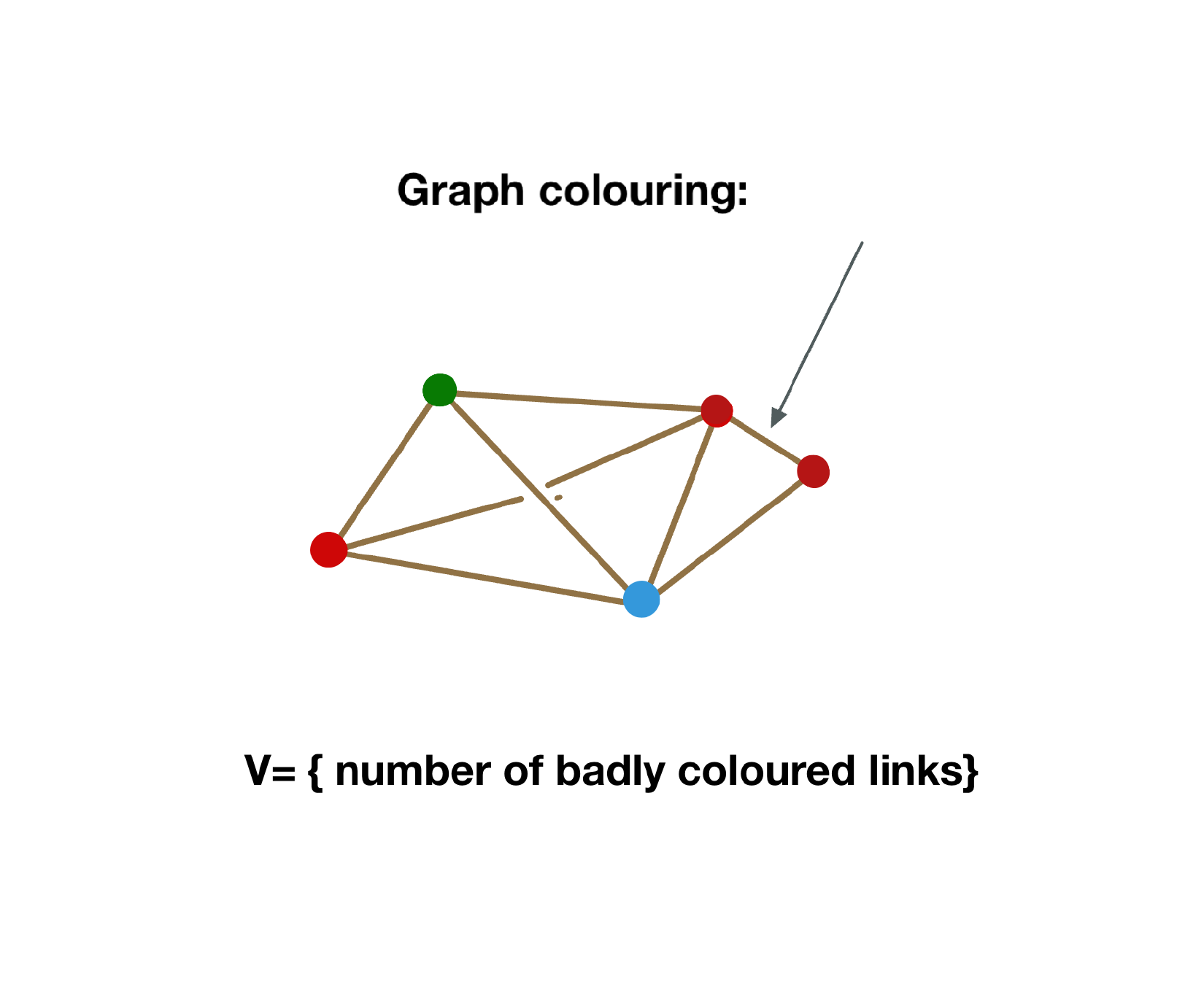}
\vspace{-1cm}
\caption{The Graph Coloring problem. The energy is the number of links joining equally colored vertices.}
\label{coloring}
\end{figure}

Consider the glassy Hamiltonians \cite{Crisanti1992,cavagna2009supercooled,anous2021quantum}:
\begin{equation}
E_p =-\frac{1}{p!} \sum_{i_1,...,i_p} J_{i_1,...,i_p} q_{i_1}... q_{i_p}
\label{pspin}
\end{equation}
where the $J_{i_1,...,i_p}$ are independent random variables with variance
$p!/2N^{p-1}$.
We often build  combinations of various $E_p$ with different $p$'s, that offer a palette of different behaviors. 
 As it stands, the potential is unbounded.  There are several ways to cure this: we may work with the $q_i$ being Ising or Potts variables, quantum or classical. For technical simplicity we choose the $q_i$ to be   classical and constrained on a sphere  $\sum_i q_i^2=N$. 
The dynamics could be anything that may be interpreted as a thermal bath
at given temperature $T_s$: quantum or classical, Monte Carlo, Glauber, Heat Bath.  Again, for technical simplicity we choose a classical Langevin process.  {\it Neither choice alters essentially the results of this paper.}
Indeed, the models we have defined contain most of what is relevant to an immense family of `hard' computational problems, such as $k$-satisfiability, and graph coloring (see Fig.  
\ref{coloring}) 

\subsection{Four versions of slow dynamics}

We consider then a system of $N$ coupled degrees of freedom $q_i(t)$ evolving by stochastic Langevin dynamics
	\begin{equation}
	\dot{q_i}(t) = -\frac{\partial V}{\partial q_i} + \eta_i(t)~,
 \label{langevin}
	\end{equation}
	where $V$ is the interaction potential, $T_s$ the (classical) temperature of the thermal bath to which the system is coupled, and $\eta_i(t)$ is a Gaussian white noise with covariance $\langle\eta_i(t)\eta_i(t')\rangle=2 T_s \delta(t-t')$.
If we deal with spherical spins, we need to impose the constraint with a time-dependent
Lagrange multiplier: $V = V_o + \lambda(t) [\sum_i q_i^2 -N]$.  Three correlations are important for the problem:
\begin{equation}
\begin{aligned}
C(t,t') &= \frac 1N \sum_i \langle q_i(t)q_i(t')\rangle ~,\\
R(t,t') &= \frac 1N \sum_i \langle q_i(t)\eta_i(t')\rangle\;\;\;\; ,\;\;\;\; \chi(t,t') \equiv \int_{t'}^t ds \; R(t,s)\\
D(t,t') &= \frac 1N \sum_i \langle \eta_i(t)\eta_i(t')\rangle - 2T_s \delta(t-t')~,
\end{aligned}
\label{CRD}
\end{equation}
The average is over the dynamic process. We shall always understand  $t\geq t'$.  For causal systems, since the noise is white,  $D=0$.

Slow dynamics happen in different possible ways, even for the same system. Consider the following:

\begin{itemize}
    \item {\bf Aging:} Starting from a high temperature configuration,
    and connecting the system with a low-temperature bath, we find that it does not manage to equilibrate in observed times, or at all. We easily see this because correlations  do not take the Time-Translational Invariant 
(TTI) form:
    $$C(t,t') = C_f(t-t')+ \tilde C(t,t') \;\;\;\; {\mbox { for example}} = C_f(t-t') + {\cal{C}}\left( \frac{t-t'}{t'}\right)$$
    so that the decay of correlation becomes slower as time passes. Furthermore,
    the Fluctuation-Dissipation Relation does not hold: 
    $$\chi(t,t') = \int_{t'}^t ds R(t,s) \neq \beta [C(t,t)-C(t,t')]$$
    \item {\bf Slowly evolving disorder:}
    Suppose that in the low temperature phase, we make the disorder evolve very slowly, stochastically,  so that:$$\langle J_{i_1,...,i_p}(t) J_{i_1,...,i_p}(t') \rangle = p!/2N^{p-1}e^{-\frac{t-t'}{\tau}}$$
   with $\tau$ large.  Aging is suppressed, because the system adapts as best it can up to the timescale $\tau$, and then keeps `chasing' the optimization  with the changing couplings. A typical correlation  may read:
    \begin{equation}    C(t,t') = C_f(t-t') + \tilde C(t,t') \;\;\; {\mbox { for example}} = C_f(t-t') + {\cal{C}}\left( \frac{t-t'}{\tau}\right) 
\label{slow}\end{equation}
 i.e. the system  becomes stationary.      However, it is {\em not} close to equilibrium. We know this because  FDR is still violated,
    as we shall see.    
    \item {\bf Weakly driven system:}
    A very similar situation to the one above arises when the system 
  has fixed disorder, but is driven by {\it non-conservative} forces: for example shear, or, in our toy models, an additional term in the equations of motion:   
  \begin{equation}
	\dot{q_i}(t) = {\mbox{ usual terms }} + \frac 1 \tau \sum_j \;A_{ij} q_j
 \label{driving}
	\end{equation}
 with the  matrix $A_{ij}$ non-symmetric, constant and dimensionless.
 Driving the system, however weakly, keeps it from aging, beyond a point.
    \item {\bf Breaking causality:}
    We may impose that the trajectory starts and ends in given configurations
  ${\bf q}_i$ and  ${\bf q}_f$, at times zero and $\tau$, respectively.
  This may force the system to  cross  barriers, for example. 
We shall also be naturally  led to considering  ${\bf q}_i={\bf q}_f$  
\end{itemize}

In the aging case, the dynamics becomes slow when the smallest time $t'$
becomes large, a self-generated large parameter. In all other cases, the chosen timescale $\tau$ is the large parameter.

\subsection{Solution for large $N$}

The solution for this problem may be obtained by summing dominant diagrams
(usually in the field theory community), or by writing a path
integral version of the dynamics and averaging (always, in the glass community). The situations above are just chosen by fixing the boundary conditions in time. Only the causality-breaking situation requires, in principle, to replicate 
the dynamics, because the normalization is no longer unity in that case.
One obtains then, in terms of   $C,R,D$  a weight of the form $e^{- N S[C,R,D]}$, which leads, after variation, to the equations of motion,
exact in the large-$N$ limit:
\begin{equation} 
\left(\frac{d}{dt} - 2T {\bf U}\right) {\bf Q} (t,t')=  \int ds\; {\bf \Sigma}(t,s) {\bf Q} (s,t') -  \mbox{\boldmath $\lambda$}(t) {\bf Q} (t,t') + \mbox{\boldmath $\delta$}(t-t')
\end{equation}
with the definitions:

\begin{gather}
\label{M}
{\bf Q}=  \begin{bmatrix}
R& C\\
D& R^t
\end{bmatrix} (t,t') \;\; , \;\;
\mbox{\boldmath $\delta$}=  \begin{bmatrix}
\delta& 0\\
0& \delta
\end{bmatrix}(t-t') \;\; , \;\;
\mbox{\boldmath $\lambda$}=  \begin{bmatrix}
\lambda & 0 \\
-\hat \lambda & \lambda 
\end{bmatrix}(t)
\;\; , \;\;
{\bf U}=  \begin{bmatrix}
0 \;& \;  1\\
0 \;  & \; 0
\end{bmatrix}
\end{gather}
\begin{gather}
{\bf \Sigma}=  p\begin{bmatrix}
(p-1) C^{p-1}R \;\;& \;\;  C^{p-1}\\
(p-1) C^{p-2} D+ (p-1)(p-2) C^{p-3}R \;\;  & \;\; (p-1) C^{p-1}R^t
\end{bmatrix} (t,t') \label{eqqq}
\end{gather}
The reader familiar with the SYK model will recognize the similarity \footnote{A notation that brings these even closer is in terms of superspace variables Eq (\ref{Q12})}.

In the glass literature, it is customary to separate explicitly the fast and the slow relaxations, which  occur at
small and large time-differences, respectively.
\begin{equation}
\begin{aligned}
\;\;\; C(t,t') &= C_f(t-t') +\tilde{{C}} \left(t,t'\right)~,\\
\;\;\; R(t,t') &= R_f(t-t') +  \tilde{{R}}  \left({t,t'}\right)~,\\
\;\;\; D(t,t') &= D_f(t-t') + \tilde {{D}} \left({t,t'} \right)~,\\
\end{aligned}
\end{equation}
If we separate the fast relaxations, for the remaining `slow' parts, the time-derivatives become small, in terms of $t'$ in
the aging case, and in terms of the parameter $\tau$ in the other three cases. In that limit, we may drop the l.h.s. of (\ref{eqqq}), and the rest  obeys a reparametrization invariance,\\
\begin{minipage}{0.4\textwidth}
\begin{equation*}
\begin{aligned}
\;\;\;\tilde C(t,t') &\rightarrow \tilde C(h(t),h(t'))~,\\
\;\;\;\tilde R(t,t') &\rightarrow \dot h(t') \;\tilde R(h(t),h(t')) ~,\\
\;\;\;\tilde D(t,t') &\rightarrow \dot h(t)\dot h(t') \; \tilde D(h(t),h(t')) 
\end{aligned}
\end{equation*}
\end{minipage} \;\;\;\;\;\;\; and \;\;\;
\begin{minipage}{0.4\textwidth}
\begin{equation}
\begin{aligned}
\lambda(t) &\rightarrow \lambda(h(t)) ~,\\
\hat \lambda (t) &\rightarrow \dot h(t) \; \hat \lambda (h(t)) 
\end{aligned}
\label{repa3}
\end{equation}
\end{minipage}~\\

Note that the anomalous dimension is zero, in this case \cite{cardy2013logarithmic}.
This pseudo invariance has an important physical meaning for glasses, as we shall see in the next section. 

\section{Physical consequences} \label{tres}

\begin{figure}
\hspace{.7in}\includegraphics[width=4.5in]{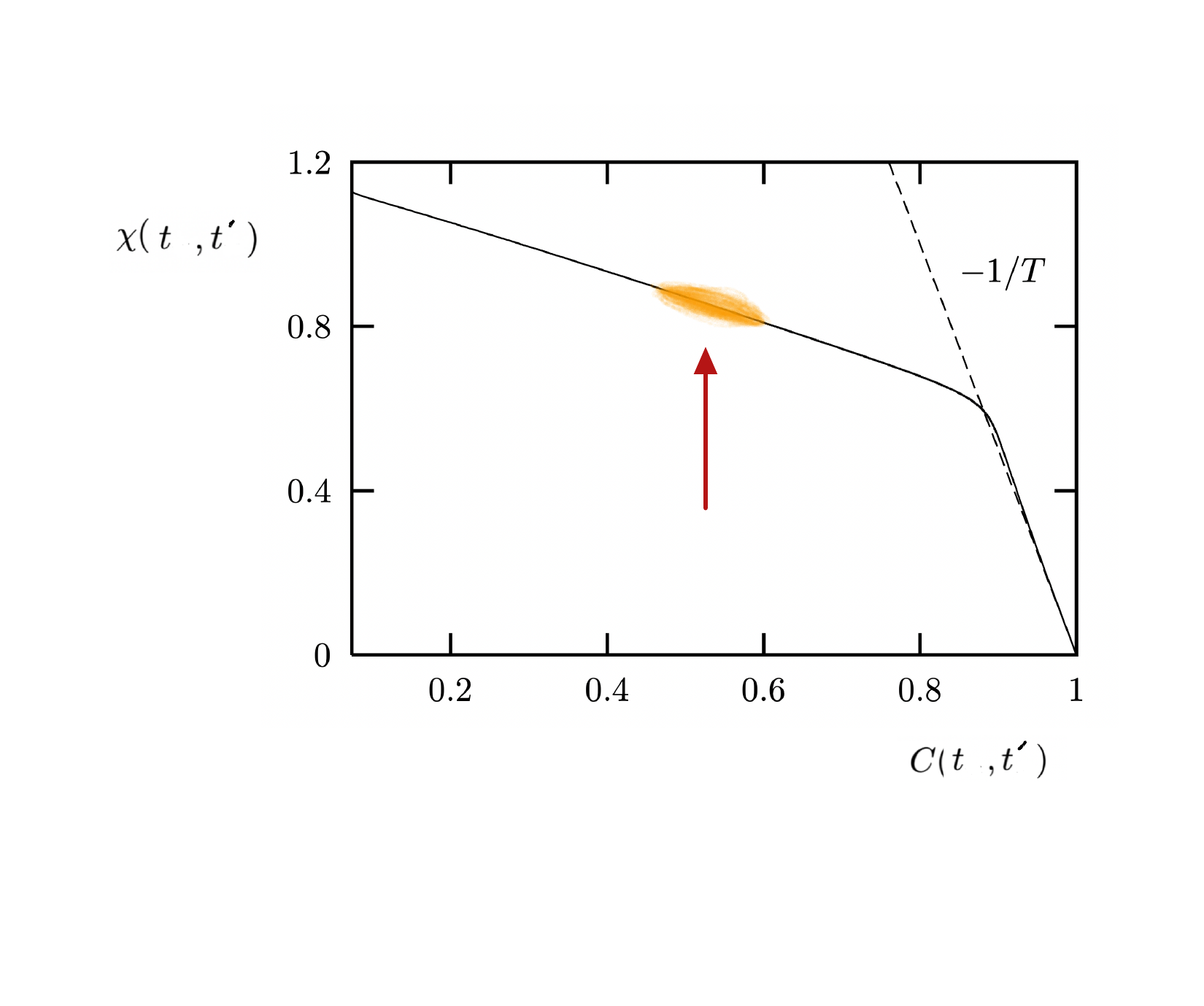}
\vspace{-2cm}
\caption{The Fluctuation-Dissipation plot breaks into two lines, one with gradient $\beta$ - as one would have in equilibrium - corresponding to fast relaxations, and one with a self-generated  $-\beta_{\mathrm{eff}}$, corresponding to slow relaxations (see Ref. \cite{CugliandoloKurchanPeliti}). Thermal fluctuations are large along the line, small in transverse directions.}
\label{aba:b}
\end{figure}

In the aging case, if one makes a parametric plot of integrated response $\chi(t,t')$ versus $C(t,t')$, one finds that for large $t'$ the curves collapse into one, sketched in Fig. {\ref{aba:b}}. If the Fluctuation-Dissipation relation were to hold, one would obtain a straight line with gradient $-\beta$. But we are out of equilibrium here, so the FDR does not apply. The result is however surprising: one obtains {\it two} straight lines, one - corresponding to small time-differences - is the one we would have obtained in equilibrium, while the other has the form we would have at another temperature $\beta_{\mathrm{eff}}$ , self-generated by the system. The emergence of an effective temperature $\beta_{\rm{eff}}$  will be further discussed  in Section \ref{siete}.

\begin{figure}
\hspace{1.5cm} \includegraphics[width=3.in]{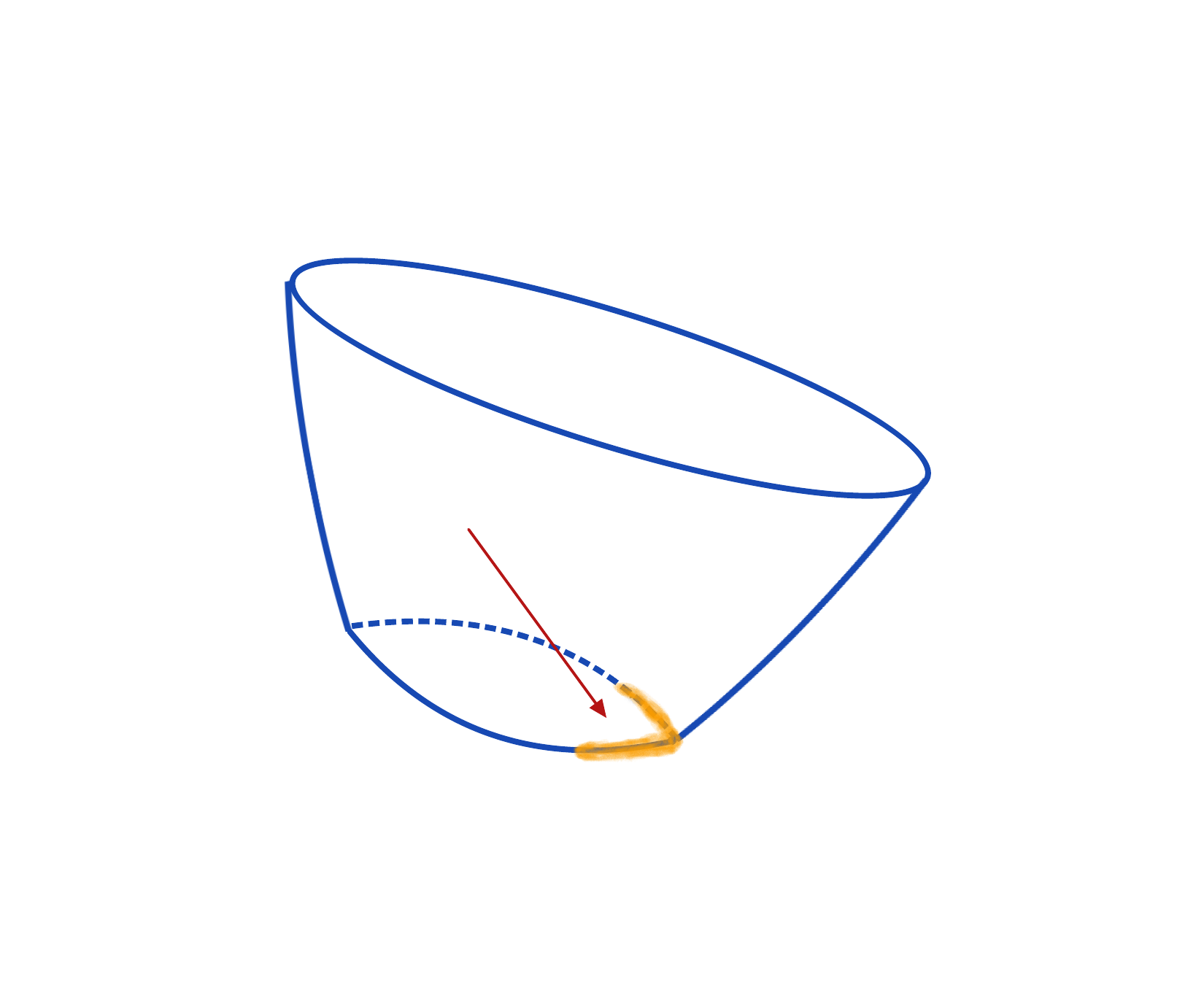}
\vspace{-1cm}
\caption{A quasi-symmetry. The thermal fluctuations are large along the `angles'.}
\label{aba:a}
\end{figure}

Next, we consider the cases of  a slowly evolving disorder,
or with weak drive. In both these situations, the system becomes stationary (only dependent on time-differences), just like a system in equilibrium. What happens with the $\chi$ vs. $C$ plot then?
Quite surprisingly, it is  identical to the one for the aging case, underlining the fact that we are stationary but still out of equilibrium.  
What has happened is that both the slow evolution of disorder, and the weak drive, {\it have dramatically reparametrized the time, changing completely the form of the slow part of the relaxation.} We should indeed not be surprised about this fact, since
time-reparametization is almost an invariance. The long time limit of  
$\chi(C)$ is instead a reparametrization invariant object, and
stays stable.
The situation is as in Fig. \ref{aba:a}: time-dependencies are
like the angle in this simple example, while reparametrization invariant quantities are like the radius.


\begin{figure}
\hspace{.2in}\includegraphics[width=3.in]{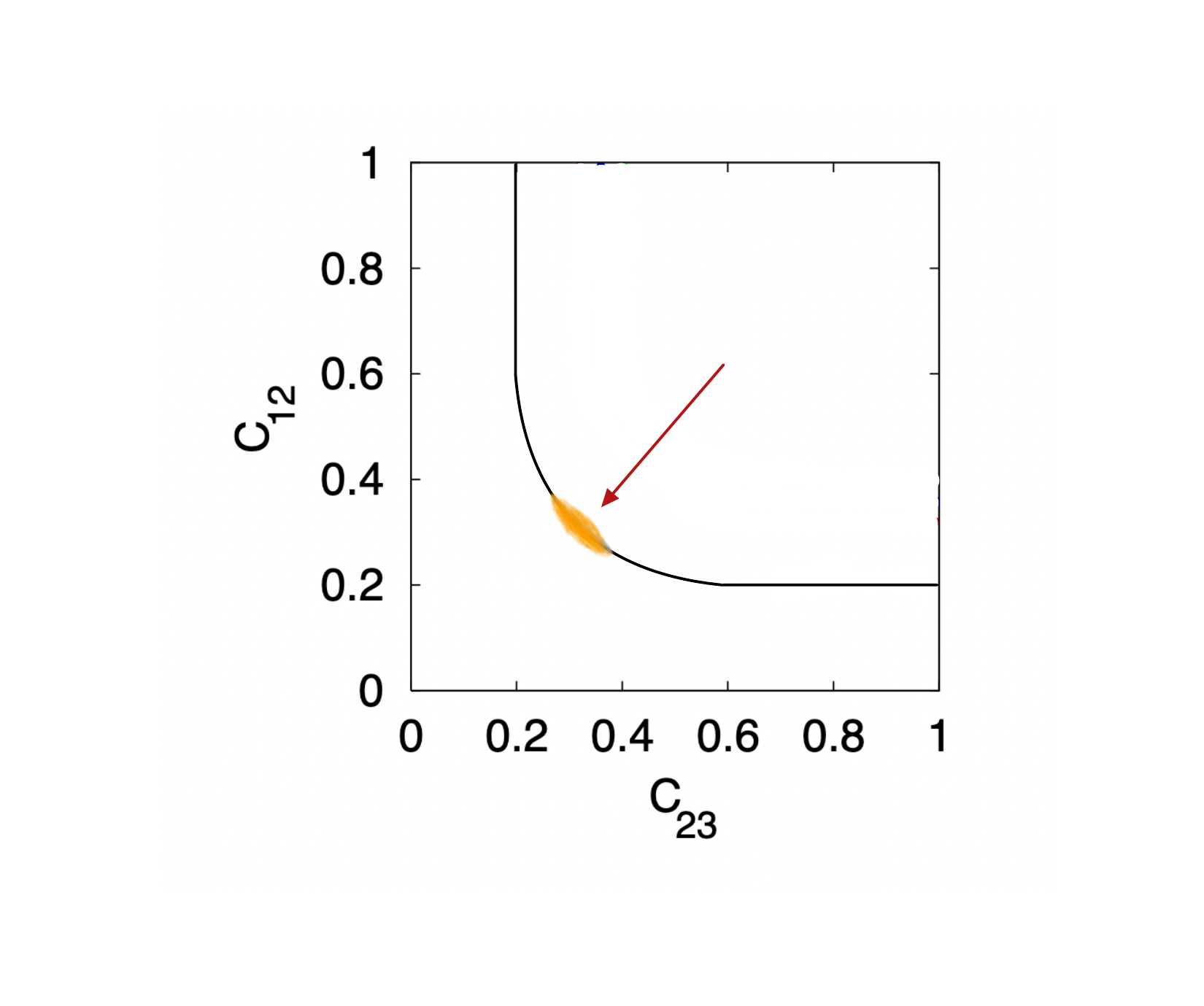}
\caption{A reparametrization invariant plot involving three correlations. In orange, a sketch of fluctuations of different realizations.}
\label{aba:c}
\end{figure}

Another reparametrization invariant quantity is constructed as follows \cite{cugliandolo1994out}: {\it i)} choose a large time $t'$ and a subsequent time
$t$ such that the correlation $C(t,t')$ takes a small fixed value, say $C=.1$. {\it ii)} For all $t'<s<t$, plot  $C(t,s)$ versus $C(s,t')$. {\it iii)} repeat for larger $t'$ (in aging, or larger $\tau$ in the other cases)
Again, this plot is the same whatever the precise time-reparametrization (see Fig. \ref{aba:c}).

In the series of papers mentioned above\cite{Castillo2002,Castillo2003,Chamon2002,Chamon2007}, the authors studied 
how spontaneous fluctuations of realistic systems are seen in a reparametrization-invariant plot such as Fig \ref{aba:b} and Fig \ref{aba:c}, and indeed found that they tend to concentrate along the lines, as shown in orange (artist view) in these two figures: indeed, an experimental version of Fig. \ref{aba:a}.

Reparametrization invariance also plays a crucial role in the activated processes in glasses. In a remarkable paper \cite{rizzo2021path}, Rizzo considered the situation in which initial and final states are deep, equilibrium states, and
there is a passage from one to the other in time $T$. Interestingly, the `slow' parts of the solutions
for given  passage times $T$, are time-reparametrizations of one another
\begin{equation}
\begin{aligned}
C(t,t') &= C_f(t-t', t/T) +{\cal{C}} \left(\frac{t-t'}{T}\right)~,~\\
R(t,t') &=  R_f(t-t', t/T) + \frac{1}{T} {\cal{R}}  \left(\frac{t-t'}{T}\right)~,\\
D(t,t') &= D_f(t-t',t/T) + \frac{1}{T^2} {\cal{D}} \left(\frac{t-t'}{T}\right)~,\\
\lambda(t) &= \ell \left(\frac{t}{T}\right)~\;\;\; , \;\; 
\hat\lambda(t) = \frac{1}{T} \; \hat \ell \left(\frac{t}{T} \right)~,
\end{aligned}
\end{equation}
The consequences of this finding have not yet been completely fleshed  out.

\section{SYK} \label{cuatro}

Here I will be brief, as the subject is treated in Sachdev's contribution in the same issue.
The Hamiltonian of the Sachdev--Ye--Kitaev model reads
\begin{equation}
H_{syk} =  {({\mathrm{i}})^{\frac{q}{2}}} \sum_{1\leq i_1<\dots<i_q\leq N} J_{i_1...i_q} \chi_{i_1} ...  \chi_{i_q}~,
\label{eq:SYK-H}
\end{equation}
where $\chi_i$ are $N$ Majorana fermions. The couplings $J_{i_1...i_q}$ are independent, identically distributed Gaussian random variables with zero mean and variance $N^{1-q} J^2 (q-1)!$.

In order to study the thermodynamics, one has to compute
\begin{equation}
\overline {\ln  Z}= \overline{\ln  {\mathrm{tr}} \left[  e^{-\beta H_{syk}} \right]}~,
\end{equation} 
where the overbar denotes an average over the couplings. This is usually done by replicating the system $n$ times and then continuing to $n\rightarrow 0$.
It turns out, however, that due to the Grassmannian nature of the degrees of freedom and the lack of a glass transition, order parameters coupling different replicas vanish, and the result coincides with the annealed average,
\begin{equation}
 \ln \overline Z= \ln \overline{ {\mathrm{tr}} \left[ e^{-\beta H_{syk}} \right]}~.
 \label{annealed}
\end{equation} 
Following precisely the same steps as in the mean-field glass, we obtain
a result exact in the large $N$ limit 
\begin{equation} 
 \ln \overline{Z}= \ln \int D[G] e^{-N S[G]}~,
\end{equation}
in terms of the correlation:
\begin{equation} \label{GG}
G(t-t') =  \sum_i \langle T \chi_i (t) \chi_i (t')\rangle~.
\end{equation} 
The large $N$ limit allows for a saddle point evaluation, and one finds that the saddle-point value of $G$ satisfies the following equation
\begin{equation} \label{eq:SYK-eom}
{ \frac{\partial G(t_1,t_2)}{\partial{t_1}}} = J^2 \int_0^\beta d t\ G(t, t_2)^{q-1}\; G(t_1, t) \  + \delta(t_1 - t_2)~.
\end{equation}
The reader familiar with Dynamical Mean Field solutions, will recognize the similarity.

The large $t$ solution of~\eqref{eq:SYK-eom}, for $T \rightarrow 0$, is ~\cite{Parcollet1999,KitaevLectures,Maldacena2016SYK,Polchinski2016,Jevicki2016}
\begin{equation} \label{eq:SYK-G0}
G(t_1,t_2) \propto \frac{1}{|t_1-t_2|^{{2}/{q}}} {\mathrm{sign}}(t_1-t_2),
\end{equation}
The effect of a a small, positive  temperature, is to cut off the  critical power law behavior, at a timescale of the order of $\beta$. 
The specific heat is linear at low temperature and, more important for us here, the model displays a positive zero-temperature entropy. 

Just as in the glassy case, the equation~\eqref{eq:SYK-eom} has, to the extent that we may neglect the time-derivative term, the approximate reparametrization invariance:
\begin{equation}
G(t_1, t_2) \rightarrow  |\dot h(t_1) \dot h (t_2)|^{1/q} G(h(t_1), h(t_2))~.
\label{repa-syk}
\end{equation}

Of course, only one specific parametrization corresponds to the true minimum of the action.
As noted by Parcollet and Georges~\cite{Parcollet1999}, 
one can obtain the low-temperature behavior  from  the zero-temperature one through a reparametrization:
 $t_a \rightarrow  \tan \left(\frac{\pi t_a}{\beta}\right)$
 ($a=1,2$)  which maps~\eqref{eq:SYK-G0} into a {\em time-translational invariant}  function of period $\beta$,
\begin{equation}
G_{\beta}(t_1-t_2) = b \left[\frac{\pi}{\beta  
\sin\frac{\pi (t_1-t_2)}{\beta}}\right]^{2/q} {\mathrm{sign}}(t_1-t_2),
\label{eq:SYK-Gbeta}
\end{equation}

The breaking of reparametrization invariance, which is a continuous symmetry, leads to the emergence of almost-soft modes governing low temperature fluctuations. 
An effective theory for reparametrizations  was constructed explicitely \cite{Maldacena2016SYK}: it allows one to compute all the low-energy properties including the main critical fluctuations, which corresponds to four-point functions, 
in particular those related to the quantum Lyapunov exponent---extracted from the so called out-of-time-order correlation function (OTOC).  This is Kitaev's intuition \cite{KitaevLectures}: the theory based on $S[h]$
has many of the symptoms one expects from a toy version of quantum gravity.

\section{Bringing together the two situations} \label{cinco}

We wish to establish a relation between glassy dynamics (classical or quantal),
with the quantum SYK model, in both of which the main role is played by an emerging reparametrization quasi-invariance. 
{\it The connection we discuss here  is not made through quantizing the original glass model, since reparametrization  is for glasses already present in classical dynamics. Indeed, it comes with a twist:}

	The evolution of the probability density following equation (\ref{langevin}) is generated by the
	Fokker--Planck operator $H_{\textrm{FP}}$,
	\begin{equation}
	\partial_t P_t(\mathbf{q}) = \sum_i \frac{\partial}{\partial q_i}
	\left[T_s \frac{\partial}{\partial q_i} + \frac{\partial V}{\partial q_i}\right] P_t(\mathbf{q}) \equiv - H_{\textrm{FP}}P_t(\mathbf{q}) .
 \label{FP}
	\end{equation}
	
	The Fokker--Planck operator is not Hermitian, but detailed balance is satisfied
	with the Gibbs distribution
	\begin{equation}
	e^{V/T_s} H_{\textrm{FP}} e^{-V/T_s}=H_{\textrm{FP}}^\dag~.
	 \end{equation}
	This allows  us to write it in an explicitly Hermitian form~\cite{Zinn-Justin_Book,Kurchan09-lectures}. Rescaling time,
	one can define the operator
	\begin{equation}\label{eq:FP-to-H}
	H = \frac{T_s}{2} e^{V/2T_s} H_{\textrm{FP}} e^{-V/2T_s} = \sum_i\left[-\frac{T_s^2}{2} \frac{\partial^2}{\partial q_i^2}
	+\frac{1}{8}\left(\frac{\partial V}{\partial q_i}\right)^2 -\frac{T_s}{4}\frac{\partial^2 V}{\partial q_i^2}\right]\ 
	\end{equation}
	$H$ has the form of a Schrodinger operator with $T_s$ playing the role of $\hbar$, unit mass and potential
	\begin{equation}\label{eq:Veff}
	V_{{\mathrm{eff}}} = \frac{1}{8} \sum_i\left(\frac{\partial V}{\partial q_i}\right)^2 -\frac{T_s}{4}\sum_i \frac{\partial^2 V}{\partial q_i^2}~.
	\end{equation}
	
	The spectrum of $H_{\textrm{FP}}$ and that of $H$ are the same, up to the rescaling in~\eqref{eq:FP-to-H}, and the 
	eigenvectors are related via the transformation above.

 The connection we establish is the following: we shall consider $H$  as our (formally)  `quantum'
 Hamiltonian. Its partition function $Z=\mathrm{tr} e^{-\beta_q H}$ is obtained by introducing a quantum temperature $\beta_q$ . Then, $T_s$ we interpret as a parameter playing the role of $\hbar$, ({\em not} the quantum temperature).
A we shall see, $H$ has a quantum phase transition at $T_q=0$ and  the value of  $T_s$  in which the original system had a dynamic glass transition $T_s=T_{\mathrm{mct}}$. 
{ What may be confusing is that, although the original system, whose partition function is $Z_c={\mathrm{tr}} \; e^{-\beta_s V}$ 
will have replica symmetry breaking at low temperature $T_s$, the `quantum' system   $Z=\mathrm{tr} \; e^{-\beta_q H}$  need not.}

\begin{figure}
\hspace{.6in}\includegraphics[width=3.5in]{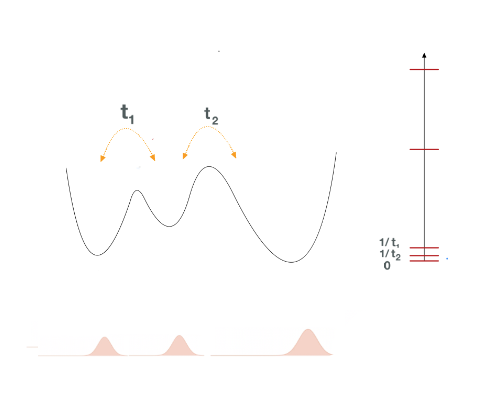}
\caption{Localized wavefunctions of the form $\langle x|a\rangle \propto e^{-\beta V }\Theta(a)$, where $\Theta(a)$ is one in the region of the basin {\bf a} and zero elsewhere. Eigenstates  of $H$ are proportional to these, eventually hybridized.
Note that as they are drawn, they are approximately orthogonal -- because they are localized -- and positive. It can be shown that the  exact `low' eigenstates may always be de-hybridized into a set of these approximately positive and orthogonal vectors.}
\label{aba:d}
\end{figure}

We can infer the spectral properties of $H$ from the general connection between spectra and  metastability \cite{Gaveau1998,Biroli2004}. Consider first a low-temperature $T_s$  Langevin process on the potential $V$ of 
Figure \ref{aba:d}.  It contains two metastable and a stable state. It is intuitive, and may be shown \cite{Gaveau1998,Bovier2002,Biroli2004}  that the spectrum of its Fokker-Planck operator has three low energy eigenstates, 
of eigenvalue the inverse of the lifetime of the corresponding metastable states in the original system - zero, if they are fully stable.
Furthermore, their wavefunctions are combinations  of the  localized, positive, distributions within each state - see Fig \ref{aba:d}. 
This is a completely general fact, and is independent of the actual origin of metastability, be it low temperature or, as in our case here, large $N$.

When we compute   $Z=\mathrm{tr} \; e^{-\beta_q H}$ we are in fact summing over trajectories following a Langevin equation (\ref{langevin}) but that, due to
a lucky realization of the noise $\eta_i(t)$, they turn out to be periodic with period $\beta_q$. 

\begin{figure}
\includegraphics[width=10cm]{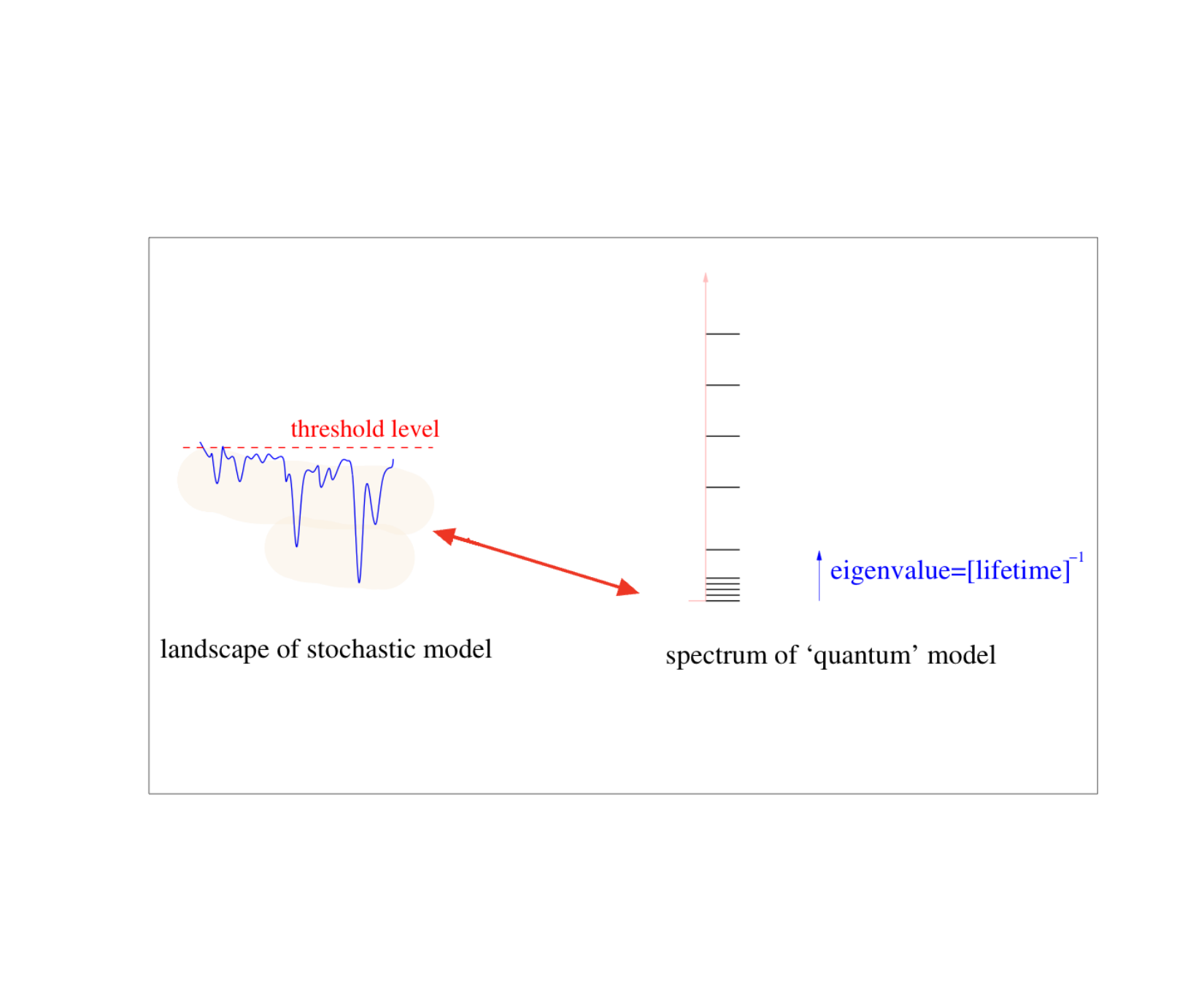}
\caption{Spectral properties of the stochastic evolution, our `quantum' Hamiltonian, and its correspondence to metastable states of the glass.}
\label{thres}
\end{figure}

\begin{figure}
\hspace{.5in}\includegraphics[angle=-90,width=4.in]{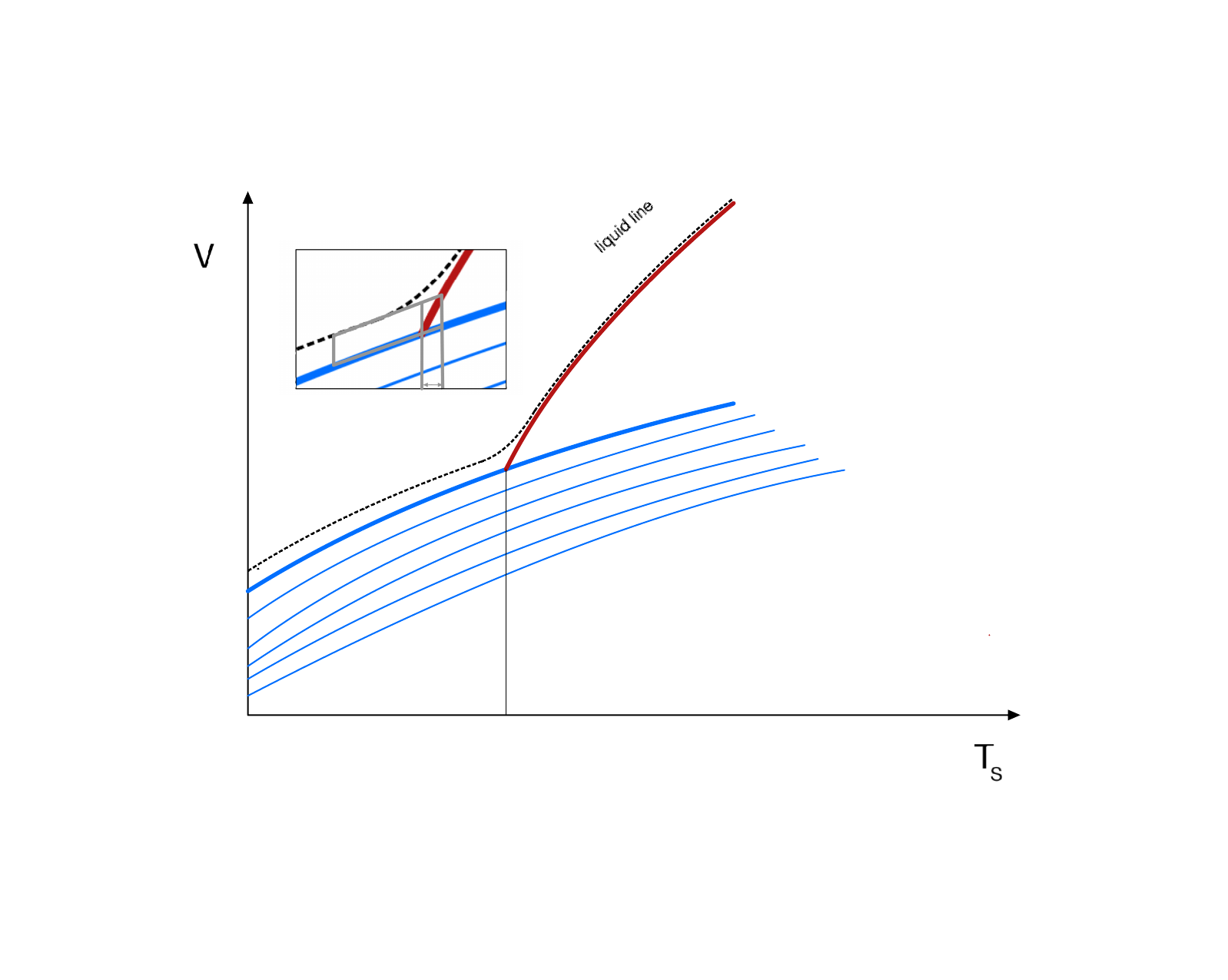}
\vspace{-1.5cm}
\caption{A sketch of the energy landsdcape as a function of the stochastic temperature. In blue, the glassy states sitting around minima of the potential, their number grows exponentially for higher  $V$, until at a certain point they become unstable. At higher $V$ the number of saddles keeps increasing, but they are no longer minima, their typical index grow with their value of $V$. The intersection with the liquid line
marks the dynamic transition $T_s$. }
\label{energy}
\end{figure}

In the case of our model glass, the situation is depicted in Figs. \ref{thres} and \ref{energy}:
We know from the theory of these models that there is an exponential number of metastable states with lifetimes 
diverging with $N$, corresponding to an exponential number of vanishing eigenvalues of $H$. Its zero $T_q$ entropy is then precisely the logarithm of this number. The argument may be generalized to barriers and the supersymmetric extension, we shall come back to this in Sect \ref{ocho}.

Suppose that we are at a stochastic temperature  $T_s$
such that the system with potential $V$ is in a glassy phase, and replica symmetry is broken when we evaluate ${\mathrm{tr}} \; e^{-\beta_s V }$
When we compute instead  $Z=\mathrm{tr} \; e^{-\beta_q H}$, we are looking for trajectories that close and are probable: for this we need that they are in the inside of states whose lifetime is not much shorter than  $\beta_q$. So we need to look for where are the most numerous of these: it turns out that they are just above the threshold level on figure \ref{thres}  
(right), see dashed line of figure \ref{energy}. For larger $\beta_q$,
trajectories have to be inside less numerous but more stable
metastable states, eventually concentrating on the threshold level of figure \ref{energy}. This situation need not necessarily be accompanied by a transition in $\beta_q$, as we shall see in section \ref{ochopuntodos}.


The direct connection is then as in Figure \ref{thres},
and holds for the dynamics of any glass model, classical or quantum (in the latter case the spectrum considered is the one of the Lindbladian):  the set of metastable states constituting the glass correspond to low eigenvalues of $H$ dominating $T_q \rightarrow 0$,
and play the same role of the states representing the `Black Hole' in SYK. The detailed  description in both cases is encoded in a theory for the reparametrizations.

Let us note that again that $H$ corresponds to supersymmetric quantum dynamics {\it restricted to
the subspace of zero fermions}. The connection between this and other supersymmetric SYK-like variants will be outlined in Section \ref{ocho}.

\section{The classical glass {\it dynamic} transition as a `quantum' critical point} \label{seis}

\subsection{The special $p=2$ case} 

In Ref \cite{facoetti2019classical}, the  $p=2$ version of the dynamics with potential Eq (\ref{pspin}) were obtained\footnote{Similar to the corresponding treatment of Ref. \cite{biggs2023supersymmetric}, but here in the zero-fermion subspace}, and are summarized in Table~\ref{tab:p2}). This case is not quite  a glass because it does not have many metastable states, but it is nice because it may be easily solved,  and it has some interesting properties. 
The original model has a transition at $T_s=T_c$. Considered as a quantum model, and evolving in real time,  at and below the transition $T_s\leq T_c$ the correlation first decays as
\begin{equation}\label{eq:p2T0asympt}
    C(t) = q + \frac{b}{t^{\frac{1}{2}}}\ .
\end{equation}
into a plateau of value $0<q<1$.
 Then it relaxes from $q$ to zero, in a timescale that diverges  as $\tau \sim \beta_q^2$ for $T_s<T_c$, but is Planckian at exactly $T_c=T_q$, i.e.  $\tau \sim \beta_q$,
and is short  when $T_s>T_c$ (see Table \ref{tab:p2}).  

From the point of view of the associated `quantum' Hamiltonian $H$, 
the specific heat displays a non-trivial scaling for $T_q\rightarrow 0$.
\begin{table}
    \caption{Summary of results for $p=2$.}\label{tab:p2}
	\renewcommand{\arraystretch}{1.6}
	{\centering
	\begin{tabular}{c|c|c|c}
	    \hline \hline
		& $T_s<T_c$ & $T_s=T_c$ & $T_s>T_c$ \\ 
		\hline 
		$q$ & $1-T_s/T_c$ & 0 & 0 \\ 
		specific heat & $ T_q^{3/2}$ & $ T_q^{3/2}$ & $e^{-  \frac{{1}}{2}(T_s-T_c)^2\;\beta_q}$ \\
		dynamics & Plateau $q+\frac{b}{t^\frac{1}{2}}$ for
		$ t\ll \beta_q^2$ & $\frac{b}{t^\frac{1}{2}}$ for
		$ t\ll \beta_q$ &
	   	$e^{- i \frac{(T_s-T_c)^2 \; t}{2T_s} }/t^{3/2}$ \\
	    \hline \hline
    \end{tabular} }
\end{table}
The vanishing of the $T_q=0$ quantum entropy is directly related to  the non-exponential number
of metastable states in the $p=2$ model. In order to obtain a different result one has to consider classical models with a much rougher energy landscape. This is what we do in the following focusing on $p>2$.

\subsection{Glassy systems with $p>2$ terms}

The classical  system of energy  given by   (\ref{pspin}) has a transition at the classical temperature $T_s=T_{\mathrm{mct}}$. 
The organization of states is sketched in Figs \ref{thres} and \ref{energy} (for $p>2$). At low energies $V$ there are many metastable states whose lifetime is exponential in $N$. At given $T_s$ their number increases exponentially with $V$, up to a point above which they lose their stability, i.e. their lifetime $\tau(V,T_s)$ becomes finite . The higher the $V$, the shorter the lifetime, but the number of
such states keeps increasing exponentially.
For  $T_s>T_{\mathrm{mct}}$ the liquid phase dominates the dynamics (the liquid line in  Fig \ref{energy}), but many metastable states also
survive above $T_{\mathrm{mct}}$ for an additional temperature interval: these are not relevant for the dynamics. The actual dynamics 
we are interested in follows the dashed line in Fig. \ref{energy}, it is in equilibrium above $T_{\mathrm{mct}}$ and out of equilibrium below.

In order to follow the physical situation, and to get rid of ambiguities, we shall consider the (slightly) modified 
Hamiltonian:
\begin{equation}
    H_\mu = H + \mu V
\end{equation}
The extra term reweights every state with a additive small quantity $ = \mu \langle V \rangle_{state}$, which guarantees that 
${\mathrm{tr}} e^{-\beta_q H_\mu}$ is dominated
by the dashed line level in Fig \ref{energy}. 
Considered as a `quantum' system, the  partition function ${\mathrm{tr}} \; e^{-\beta_q H_\mu}$ has a 
`quantum' transition only in the limit $\mu \rightarrow 0$,  between a situation with a finite entropy at $T_q=0$ when  the parameter $T_s<T_{\mathrm{mct}}$ and zero  when $T_s>T_{\mathrm{mct}}$. The transition point is then ($T_s=T_{\mathrm{mct}}$  , $T_q=0$, $\mu \rightarrow 0$).
This point corresponds to the so-called `Mode Coupling Transition' of glass theory \cite{reichman2005mode,GotzeLesHouches}, hence the notation $T_{\mathrm{mct}}$.

The complete solution of this problem is yet to be done, see Refs \cite{Biroli2004,facoetti2019classical} for partial results.
A rather crude argument (see inset in Fig \ref{energy}) gives interesting hints: consider a situation in which $T_s$ is just below $T_{\mathrm{mct}}$, and we fix through 
 a small value of $\mu$ the  value ${\cal{V}}_\mu = \langle V \rangle_\mu $. Then
 $\varepsilon \equiv  ( {\cal{V}}_\mu - {\cal{V}}_0 ) $ is correspondingly small.
The main assumption is  to estimate the timescale $\tau$ of the correlation decay as being the same as the one in equilibrium 
at the temperature $T_s>T_{\mathrm{mct}}$, such that the liquid energy is $\varepsilon$ above the highest glass energy, the argument being that the metastable states contributing to the dynamics are the same. From Mode Coupling Theory, we know
that $ \tau \propto (T_s-T_{\mathrm{mct}})^{-\gamma}$ where $\gamma$ is a mode-coupling exponent.
The extensive eigenstates of $H_\mu$ contributing to this are of order $\epsilon \sim \tau^{-1}$ per degree of freedom\footnote{That the eigenstate of the global system should scale with the inverse timescale times the size may be 
understood easily in the case of uncoupled systems}, while the log of the number of metastable states of this value is just linear in $S \propto \beta_{\mathrm{eff}} \; \varepsilon  $, where $\beta_{\mathrm{eff}}$ is a constant. Again $T_s-T_{\mathrm{mct}}$ and
$\varepsilon$ are proportional, so $\tau \propto \varepsilon^{-\gamma}$.
With these assumptions, we may determine $\varepsilon$ in terms of the saddle point
in the exponent
$Z = Z_0 \; e^{N[ \beta_{\mathrm{eff}} \varepsilon- b \beta_q \varepsilon^{-\gamma}] }$, to obtain
$
 T_q   =T_{\mathrm{eff}} b \gamma \varepsilon ^{\gamma-1}
$. 
 From this, we get an estimate of the timescale:
 \begin{equation}
     \tau \propto \left(\frac{\beta_q}{\beta_{\mathrm{eff}}}\right)^{\frac{\gamma}{\gamma-1}}
 \end{equation}
      Remarkably, $\gamma$ is known to be $1 \le \gamma\le \infty$ \cite{GotzeBook,GotzeLesHouches}. In fact, one may tune it to any value in this range, just by playing with combinations of the terms $E_p$ in (\ref{pspin}).  This means that the 
 timescale is always longer than Planckian. This suggests that in imaginary time, we should not expect to see
 any descent away from the plateau value $q$ (Matsubara time is never long enough). $q$ is the `size' of the state. The scale would only become Planckian as $\gamma \rightarrow \infty$: a value in the parameters that has a precise meaning: it is a triple   point  where   the transition of the original glass stops being from replica symmetric to one-step replica symmetry breaking, and becomes  replica-symmetric to full replica symmetry- breaking (see Ref. \cite{crisanti2004spherical}): the whole organization of saddle points in the landscape changes there.  

Even if this argument is tentative, it points out that something interesting may happen at that triple point of $V$.
\section{Coupling two systems: thermo-field double} \label{siete}

As mentioned above, a property of slow dynamics in a model given by Eq. (\ref{pspin}) is that when kept out of equilibrium in any of the ways described in Section \ref{dos} we obtain a 
fluctuation-dissipation plot as in Fig. \ref{aba:b}, where an effective temperature $\beta_{\mathrm{eff}}$ emerges. The reason for this was unexpected \cite{Cugliandolo_1993_Analytical,CugliandoloKurchanPeliti} and is non-trivial: the slow dynamics explores 
metastable states in a `democratic way' so that the measure is, for all practical purposes $$|TF\rangle \propto \sum_a e^{-\frac 12 \beta_{\mathrm{eff}}  F(a)}|a\rangle$$ where $F(a)$ is the free energy (at temperature $T_s$) restricted to state `$a$' of `size' $q$: 
$$q = \frac{1}{N} \sum_i \langle a|q_i|a \rangle^2 / \langle a|a \rangle^2  $$
the Edwards-Anderson parameter.
Here, as discussed  in Section \ref{cinco} (see Fig. \ref{aba:d}), each low-lying state is of the form $\langle x|a\rangle \propto e^{-\frac \beta 2 V(x) }\Theta(a)$,
where $\Theta(a)$ has support in $a$.  (The $1/2$  factor in the exponent comes from the transformation Eq. (\ref{eq:FP-to-H}), and thus corresponds to a Gibbs-Boltzmann distribution)

 Two temperatures appear: the bath one inside the state, and the other one $\beta_{\mathrm{eff}}$
that reflects the choice of the metastable  states explored by  the nonequilibrium dynamics.
\begin{figure}
\vspace{-1in}
\begin{center}
\hspace{.0in}\includegraphics[width=5.5in]{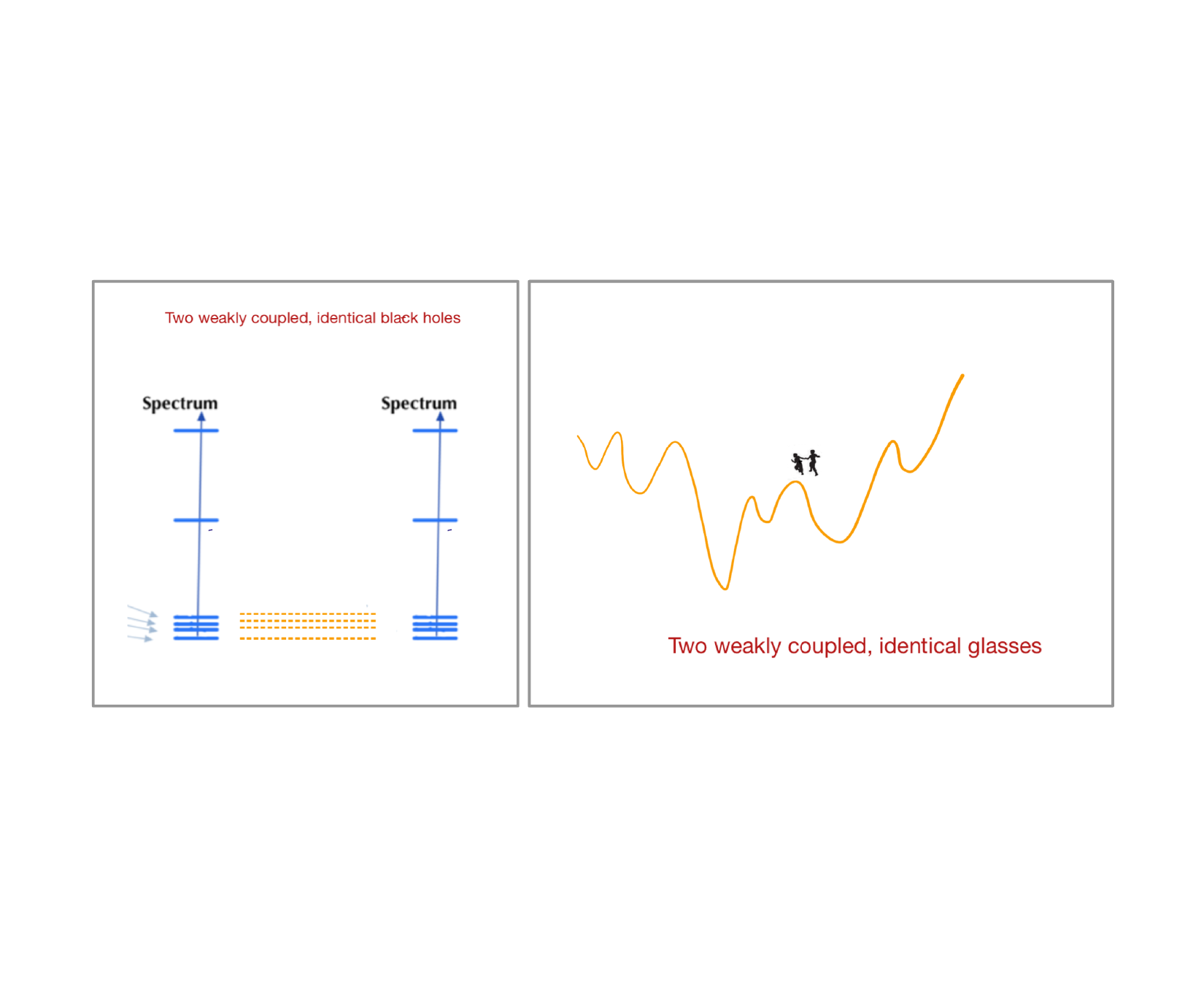}\hspace{1cm}
\end{center}
\vspace{-2cm}
\caption{Weakly coupled walkers explore the same state at each time, but independently inside it.}
\label{coupled}
\end{figure}

Consider now {\it two} systems with the same energy given by Eq. (\ref{pspin}), kept out of equilibrium in the same way as before (see Sect. 5.2 of Ref \cite{kurchan2023time}). If the systems are uncoupled, the dynamics follow their own paths, and the overlap between systems is zero. Instead, if we couple the systems just strongly enough in a way as to favor them moving near each other (Fig. \ref{coupled})
\begin{equation}
  V^{(2)} =   -\frac{1}{p!} \sum_{i_1,...,i_p} 
  J_{i_1,...,i_p} \left\{q_{i_1}... q_{i_p}+q_{i_1}'... q_{i_p}'\right\} + \epsilon \sum_i q_i q_i'
\end{equation}
the systems will be having their fast evolutions independently, {\em but at all times  within the same state}, i.e. they `jump together'. {\it For exactly the same reasons as for two coupled SYK models, in the limit of slow dynamics, the coupling needed to keep the systems in pace becomes small in the limit of large $\tau$, }  their fast motion is uncoupled,
while the slow motion is coupled -- the infrared is coupled while the ultraviolet is not.
Because metastable states are then `locked' to be 
in the same basin for both systems,
the eigenstates are then of the form
\begin{equation}
    |aa\rangle = |a\rangle_1 \otimes |a\rangle_2
\end{equation}
or hybridizations thereof.  \cite{cottrell2019build} 
Again, to a good approximation this means that the distribution is 
\begin{equation}|TFD\rangle   \propto \sum_a e^{-\beta_{\mathrm{eff}} F(a)}
|   a  a      \rangle \label{tfd} \end{equation}
so we discover that the coupled slow dynamic  is naturally represented by the thermo-field double,
in a way that is very reminiscent of the construction in Ref.  \cite{cottrell2019build}.
Let us remark in passing that making two replicas evolve together is actually a good practical method
for optimization, see Ref \cite{pittorino2021entropic})

\section{Supersymmetric models as glassy dynamics} \label{ocho}

\subsection{Hamiltonians}

As we mentioned above, $H$ may be promoted to full SUSY quantum mechanics by adding fermions:
\begin{equation}
    H_{\mathrm{susy}}= H+ \frac{T_s}{2} \sum_{ij} \frac{\partial^2 V}{\partial q_i \partial q_j} a^\dag_i a_j
    \label{susy}
\end{equation}
We may equivalently write this in the base of the Fokker-Planck operator
\begin{eqnarray}
	H_{{\mathrm{FP-susy}}} &=& - \sum_i \frac{\partial}{\partial q_i}
	\left[T_s \frac{\partial}{\partial q_i} + \frac{\partial V}{\partial q_i}\right] + \frac{T_s}{2} \sum_{ij} \frac{\partial^2 V}{\partial q_i \partial q_j} a^\dag_i a_j 
 \nonumber \\ &=&- \sum_i P_i
	\left[T_s P_i + \frac{\partial V}{\partial q_i}\right] + \frac{T_s}{2} \sum_{ij} \frac{\partial^2 V}{\partial q_i \partial q_j} a^\dag_i a_j
 \label{FP1}
	\end{eqnarray}
where $P_i \equiv  \frac{\partial} {\partial q_i}$\footnote{A notation that has been employed in spin-glass dynamics\cite{Kurchan1992susy}, is to encode the order parameters in a single superspace one
\begin{equation}
\begin{aligned}
Q(1,2)&=
C(t_1,t_2) + (\bar \theta_2 - \bar \theta_1)
\theta_2  R(t_1,t_2)
+ \bar \theta_1 \theta_1 R(t_2,t_1) + \bar\theta_1\theta_1 \bar \theta_2 \theta_2 D(t_1,t_2) \\ 
&+  \text{odd terms in the $\bar \theta,\theta$}~.
\end{aligned}
\label{Q12}
\end{equation}
It is useful because $Q$ plays a role that is formally very similar with the replica matrix $Q_{ab}$ in the limit $n \rightarrow 0$. I will not use it here, but the reader may find details in the appendix A of Ref \cite{facoetti2019classical}, and references therein.} .  

The zero-fermion subspace anihilated by $a_i$ represents the 
Fokker-Planck equation, which in turn derives from the Langevin Eq (\ref{langevin})\footnote{The  subspaces of higher number $k$  of fermions may be associated with another dynamics that has been used to locate saddles of index $k$~\cite{tuanase2004metastable}.}.
We may choose the fermion number $N_f=\langle \sum a^\dag_i a_i \rangle$ subspace
directly, or by adding a chemical potential $H_{\mathrm{susy}}+z N_f$ in the Hamiltonian.  If $z=0$ , typically half-filled $N_f\sim N/2$ subspaces dominate, because those saddles are overwhelmingly more numerous in a disordered system.\\
$\bullet$ Physical  glasses are about the $N_f=0$ subspace, although the theory  gives us indications on higher $N_f$ as well. \\
$\bullet$ To the best of my knowledge supersymmetric models have been studied (except for a brief mention in Ref \cite{biggs2023supersymmetric}) without fixing the fermion number, which usually amounts to being dominated by half-filling.

\begin{figure}
\hspace{.7in}\includegraphics[width=3.in]{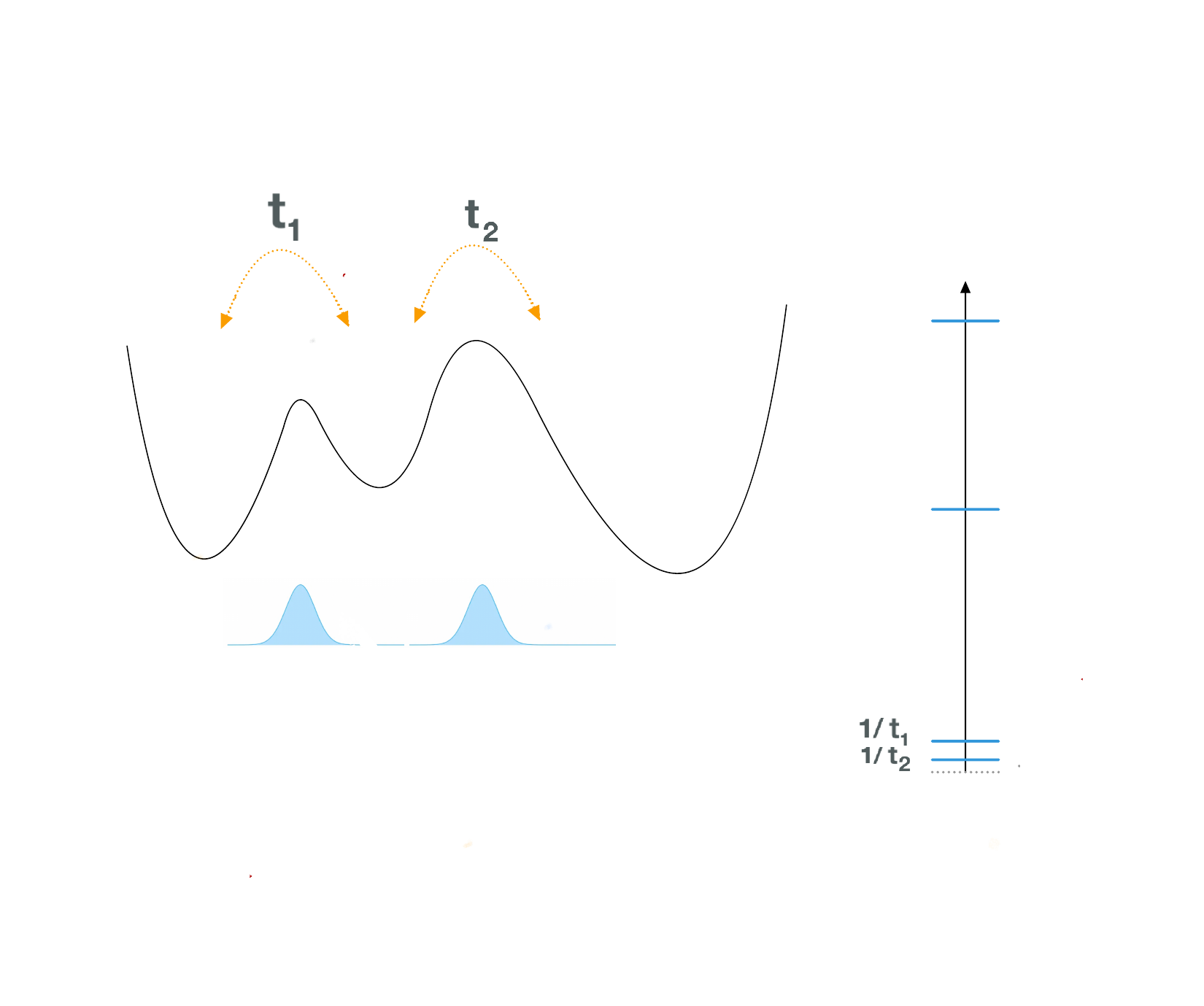}
\caption{Same as in Fig. \ref{aba:d}, but for one-fermion states of $H_{\mathrm{susy}}$. `Barrier states' are like metastable states, but with $k$ unstable directions.}
\label{aba:e}
\end{figure}

Metastable states are then generalized to the case with fermions (\ref{susy}) : low $k$-fermion eigenstates are concentrated on barriers of index $k$  (see Fig. \ref{aba:e}),  and
one may in general reproduce the construction for Morse theory \cite{Witten1982}, again {\it generalized to any possible origin of the metastability other than low temperature} \cite{tuanase2004metastable},
most notably large $N$.

Let us compare this to a family of supersymmetric models in the literature \cite{anninos2016disordered,murugan2017more} - here we follow the notation of Ref \cite{biggs2023supersymmetric}.

\vspace{.2cm}
{\bf $N=2$}
\vspace{.1cm}

\begin{eqnarray}
  S &=& S_f + S_f  \\
  S_b&=& \int d\tau \; \; \left[ \frac 12 (\dot \phi_i)^2 + \frac 12 (F^i)^2 + 3i C_{ijk} F^i \phi^j \phi^k \right] \nonumber \\
  S_f&=&  \int d\tau \; \; \left[ \delta_{\alpha \beta}\delta_{jk} \frac {\partial}{\partial \tau} +  3i C_{ijk} F^i  \right]\psi_\alpha^j \phi_\beta^k
\end{eqnarray}

Let us make the change of variables $P_i = i F_i + \dot \phi$. We obtain, up to a boundary term (i.e. we are making a canonical transformation):
\begin{equation}
    S_b= P_i \dot \phi_i - \frac 12 P_i^2 -  P_i \left(\frac{\partial^i V}{\partial \phi_i}\right)
 \end{equation}
where $V =  3 \sum C_{ijk}  \phi_i \phi^j \phi^k$
Finally, going to Dirac fermions  $a^\dag_i =\frac 1 {\sqrt 2} (\psi_i^1 + i \psi_i^2)$
we conclude the action is exactly of the form~\footnote{I have been careless with factor orderings} 
Eq (\ref{FP1}), 
and  we are dealing with a stochastic problem Eq (\ref{langevin}) at stochastic  temperature $T_s=1$ in the landscape Eq. (\ref{pspin}). Let us stress again that we know this landscape in great detail, in particular the density of saddles of each index at each level of $V$ ~\\~\\
$\bullet$ Eigenstates have eigenvectors localized around saddles of $V$ \\
$\bullet$ The Lagrange multiplier $\lambda \sum q_i^2$ 
term in $V$ is absent here, but it could be added. The potential 
would then be bounded -- it always is in the glass literature.
  In this constrained case, there is no qualitative difference between odd and even $p>2$.\\
  $\bullet$ For the constrained case, the Betti numbers (and hence the number of states anihilated by
both SUSY charges) are those of a sphere ~\\
$\bullet$ As is well known (\cite{Zinn-Justin_Book}) the $N=2$ supersymmetry has the physical meaning  that the heat bath is in  thermal equilibrium  -- the stochastic dynamics satisfy {\it detailed balance}.\\
$\bullet$ One may guess that the $T_q=0$ entropy of a model with $N_f$ fermions is equal to the number of saddles of the potential with index $N_f$. Indeed, when the fermion number is not fixed, as in Ref \cite{biggs2023supersymmetric},  the result  is $S_o=\frac 12 (p-1) N$, the known total number of {\it real saddles} \cite{kent2021complex} - the square root of the overall number of complex saddles.

\vspace{2.2cm
}
{\bf $N=4$}

\vspace{.1cm
}
The complex version of the model was introduced by Anninos et al \cite{anninos2016disordered}.
\begin{eqnarray}
  S &=& S_f + S_f  \\
  S_b&=& \int d\tau \; \; \left[ \frac 12 \dot{\bar{\phi}} \dot \phi_i + \frac 12 \bar F_i F^i + {\mathrm{Re}} \left\{ 3i C_{ijk} F^i \phi^j \phi^k \right] \right\} \nonumber \\
  S_f&=&  \int d\tau \; \; {\mathrm{Re}} \left[ \delta_{\alpha \beta}\delta_{jk} \frac {\partial}{\partial \tau} + {\mathrm{Re}} \left\{  3i C_{ijk} F^i  \right] \right\} \bar\psi_\alpha^j \phi_\beta^k
\end{eqnarray}
Using the Cauchy-Riemann equations, and repeating the same changes of variables as above, one concludes that the problem has now  $2N$ variables $(\phi^i_R,\phi^i_I)$ and a landscape ${\mathrm{Re}}\;  V (\phi^i_R,\phi^i_I)$. \\
$\bullet$ From Bezout's theorem, the number of saddles (which are isolated) is exactly $(p-1)^N$. Because each saddle corresponds to a low-lying eigenstate of $H_{\mathrm{susy}}$ (with $N$ fermions), the zero-temperature entropy is $(p-1)\ln N$. Both results were found  in 
Ref \cite{biggs2023supersymmetric}.~\\
$\bullet$ From the Cauchy-Riemann equations, we find that all saddles have index exactly  equal to $N$ (half of the directions descend).
The gradient lines span a  `Lefschetz thimble', leading to a saddle. ~\\
$\bullet$ The organization of saddles of this  landscape was studied in Re. \cite{kent2021complex}, albeit within the replica-symmetric ansatz.

\subsection{Replica symmetry breaking} \label{ochopuntodos}

As we have done above,
given the structure of supersymmetry, we may expect that the large  $\beta_q$ behavior (with, in addition, $T_s  \rightarrow 0$ in the  spherically constrained case) may be inferred from the statistics of saddles of each index. To determine the statistics of saddles of each index, one can use an approach that has a long tradition in spin glasses \cite{Bray_1980_Metastable}, in the physics \cite{Cavagna_1997_An,Cavagna_1997_Structure,Cavagna1998,Cavagna_1998_Stationary,Cavagna_2005_Cavity,Fyodorov_2004_Complexity,Fyodorov_2007_Replica,Ros2018,Ros_2019_Complexity,kent2021complex} and, subsequently,  in the mathematics  \cite{Auffinger_2012_Random,Auffinger_2013_Complexity,BenArous_2019_Geometry} literature.
One starts from the `time-less'  version of
equation (\ref{pspin}) and sets up a `Kac-Rice' (aka Fadeev-Popov) integral \cite{Kac_1943_On}
(see also Ref \cite{lin2023holography}):
\begin{equation}
  I=  \int \Pi_i \; d 
  \eta_i \; dq_i \;\; \delta\left(\frac{\partial V}{\partial q_i} + \eta_i \right) \; \; \det \left| \frac{ \partial^2V}{\partial q_i q_j }\right| e^{- T_s\eta_i^2}
\end{equation}
 A timeless noise $\eta_i$ is added for convenience.
The trick in this context is to realize \cite{Fyodorov_2004_Complexity,Fyodorov_2007_Replica,Bray_2007_Statistics} that  the second derivatives and the argument of the delta are statistically independent (a property of Gaussian distributions), and may be computed separately. The index of saddles may thus be fixed
at the end of the calculation, without any need of fermions.  One performs the same steps as in dynamics, introducing:
\begin{equation}
\begin{aligned}
C^{ab} &= \frac 1N \sum_i \langle q_i^a q_i^b\rangle ~,\\
R^{ab} &= \frac 1N \sum_i \langle q_i^a\eta_i^b\rangle\;\;\;\; \\
D^{ab} &= \frac 1N \sum_i \langle \eta_i^a\eta_i^b\rangle - 2T_s ~,
\end{aligned}
\label{CRD1}
\end{equation}
This time we have replicated the system $n$ times, as we should have done for the dynamics, since we need to compute the average of  $\ln  I$.
Recently \cite{kent2023count}, the full replica symmmetry breaking ansatz for this Kac-Rice approach
was obtained.
It confirms the lack of replica symmetry breaking for the parameters  corresponding to
 the $N=2$ model of Ref \cite{biggs2023supersymmetric}.
The same method was used to compute the saddles in the complex case \cite{kent2021complex} (but only withing
replica-symmetry), 
but one does not expect RSB for the associated SUSY model either. In both cases the reason is  the nature of the saddles
that dominate.  

It is  useful to recall that a `pure state' in the sense of the Parisi ansatz is in general composed of many saddles. What separates the sets of saddles that belong to two different states are the clustering properties, see M\'ezard et al \cite{MezardParisiVirasoro}.

\section{Perspectives} \label{perspectives}

There are many interesting lines to pursue. Here I  arbitrarily choose two.

From the point of view of glasses, it would be interesting to understand what are the  conformal properties of the `infrared' theory describing reparametrizations.  Some of this may be already learnt   form the treatment of supersymmetric models in Refs \cite{anninos2016disordered,Murugan2017,biggs2023supersymmetric}, adapted to the zero-fermion subspace. Another interesting lead might be to apply the ideas in Ref. \cite{cardy2013logarithmic}, where $c=0$  conformal theories arising from replicated disordered models are discussed.

From the point of view of quantum models, perhaps more levels of the dynamic or replica {\it ultrametric} hierarchies \cite{mezard1987spin}  may turn out to be relevant. Consider the slow dynamics as described Equation (\ref{slow}). Depending on the relative weight of different terms in a sum of $E_p$ in (\ref{pspin}), one can construct models
with two or more hierarchically nested timescales, e.g:
\begin{equation}    C(t,t') =  C_f(t-t') + {\cal{C}}_1\left( \frac{t-t'}{\tau^a}\right) 
+ {\cal{C}}_2\left( \frac{t-t'}{\tau^b}\right) 
\label{slow1}\end{equation}
The timescales are nested for large $\tau$ if $a<b$. Even infinitely many nested timescales are possible --
this is what happens with the Sherrington-Kirkpatrick model:
\begin{equation}    C(t,t') =  C_f(t-t') + {\cal{C}}\left( \frac{\ln(t-t'+1)}{\ln \tau}\right) 
\label{slow2}\end{equation}
where the times are adimensionalized, see Ref \cite{contucci2021stationarization}.
As mentioned in section, the parameter $\tau$  may be introduced as  $H_{\mathrm{susy}} + \frac{1}{\tau} H_{\mathrm{drive}}$, where $H_{\mathrm{drive}} = \sum_{ij} [P_i A_{ij} q_j + a^\dag_i a_j A_{ij}]$ is the (non Hermitean) operator corresponding to nonconservative forces Eq (\ref{driving}) through Eq (\ref{FP1}) ,
and do first the limit $\beta_q \rightarrow \infty$, and only then $\tau$ large. 
Perhaps the same result is obtained  with $H_{\mathrm{susy}} + \frac{1}{\tau} V$, although this has not
been done yet.
In such a situation, one may couple two systems  with an interaction that is strong with respect to $\frac {1}{\tau^a}$ and weak with respect to $\frac {1}{\tau^b}$ or one which is strong with respect to both. The pair of systems admit then two different types of `wormholes', either coupling the systems for correlations corresponding to ${\cal{C}}_2$ or for those corresponding to  ${\cal{C}}_1$ and ${\cal{C}}_2$.

\begin{figure}
\hspace{.9in}\includegraphics[width=2.9in]{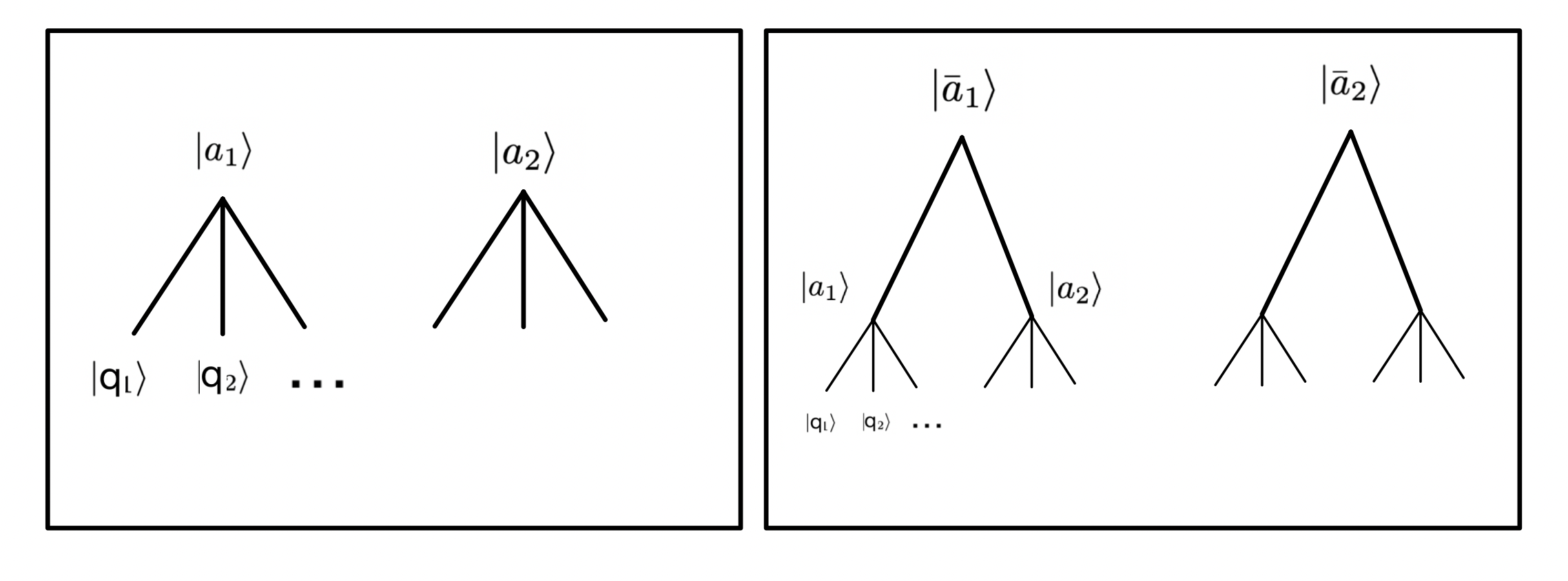}\\
\hspace{2cm}\includegraphics[width=12cm]{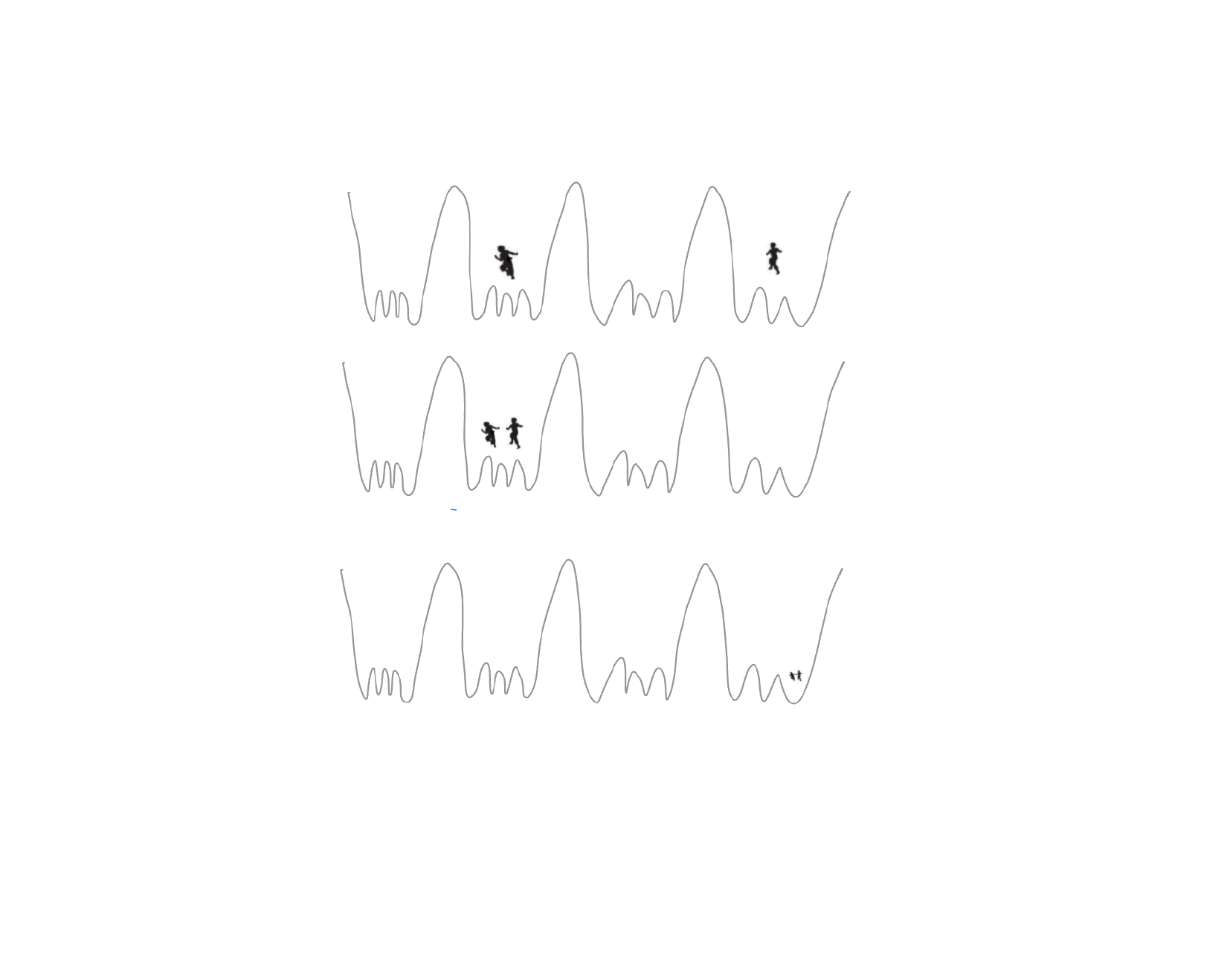}
\vspace{.5cm}
\caption{Hierarchical construction of states}
\label{ultra}
\end{figure}

This suggests that  there may be thermo-field doubles of two kinds, based on the states $|a\rangle$ as in Eq (\ref{tfd}), and also based on states $|\bar a\rangle = \sum_{a \in \bar a}  e^{-\frac 12 \beta_{\mathrm{eff}}^1 F(a) }|a>$  and then constructing
$\sum_{\bar a} e^{-\beta_{\mathrm eff}^2 F_{\bar a} } |\bar a\bar a\rangle $ as in Fig. \ref{ultra}.
I do not know if this has been done, or if it is relevant for more general wavefunctions. The scheme may be repeated 
to any number of levels.

\section{Conclusions}

If the connections pointed out in this paper turn out to be  fruitful, the suggestion would be
that glass physicists may be curious to take a deeper plunge in
reparametrization and conformal questions, and those interested in  toy gravity models, in replica theory and glassy dynamics. 

\section*{Acknowledgments}

I wish to thank G. Biroli, L. Cugliandolo, J. Jacobsen, M. Picco and  D. Reichman for useful discussions.

\bibliographystyle{ws-procs961x669}
\bibliography{Solvay27_procs.bib}

\end{document}